\documentclass[preprint,review,10pt]{elsarticle}
\usepackage{graphics}
\usepackage{graphicx}
\usepackage{epsfig}

\usepackage{subfigure}
\usepackage{tikz}
\usepackage{lineno}
\usepackage{amssymb}
\usepackage{amsmath}
\usepackage{epsfig}
\usepackage{rotating}
\usepackage[hang,stable]{footmisc}
\usepackage{color}
\usepackage{bm}
\usepackage{multirow}
\usepackage{dcolumn}
\usepackage{booktabs}
\usepackage{ifpdf}
\usepackage{amsthm}
\usepackage{CJK}

\usepackage{setspace}
\usepackage{subfigure}
\usepackage{float}
\usepackage{graphicx}
\usepackage{ listings}
\usepackage[lined,boxed,commentsnumbered, ruled]{algorithm2e}
\newtheorem{remark}{Remark}
\newtheorem*{problem0}{Problem P}

\newtheorem*{problemN}{Problem P$_N$}

\newcounter{test}
\journal{Asia-Pacific Journal of Chemical Engineering}
\begin{document}
\begin{CJK*}{GBK}{kai}
\begin{frontmatter}


\title{Dynamic Optimization of Trajectory for Ramp-up Current Profile in Tokamak Plasmas\footnote{This work was supported by the National Natural Science Foundation of China grant 61473253, the National High Technology Research and Development Program of China (2012AA041701), and supported by Collaborative Innovation Center for Industrial Cyber-Physical System and the Fundamental Research Funds for the Central Universities (2016XZZX002-01)}}


\author{
Zhigang Ren$^{a}$, Chao Xu$^{b,}$\footnote{Correspondence to: Chao Xu, Email: cxu@zju.edu.cn}, Yongsheng Ou$^c$}
\address{$^a$ Laboratory of Information \& Control Technology, Ningbo Institute of Technology, Zhejiang University, Ningbo, China.\\
$^b$ State Key Laboratory of Industrial Control Technology and Institute of Cyber-Systems \& Control, Zhejiang University, Hangzhou, Zhejiang, China.\\
$^{c}$ Shenzhen Institutes of Advanced Technology, Chinese Academy of Sciences,
Shenzhen, Guangdong, China.\\
}


\begin{abstract}
In this paper, we consider an open-loop, finite-time, optimal control problem of attaining a specific desired current profile during the ramp-up phase by finding the best open-loop actuator input trajectories. Average density, total power, and plasma current are used as control actuators to manipulate the profile shape in tokamak plasmas. Based on the control parameterization method, we propose a numerical solution procedure directly to solve the original PDE-constrained optimization problem using gradient-based optimization techniques such as sequential quadratic programming (SQP).
This paper is aimed at proposing an effective framework for the
solution of PDE-constrained optimization problem in tokamak plasmas.
A more user-friendly and efficient graphical user interface (GUI) is designed in MATLAB and the numerical simulation results are verified to
demonstrate its applicability. In addition, the proposed framework of combining existing PDE and numerical optimization solvers to solve PDE-constrained optimization problem has the prospective to target challenge advanced control problems arising in more general chemical engineering processes.
\end{abstract}


\end{frontmatter}



\section{Introduction}
%
%
%
Nuclear fusion is two light nuclei such as deuterium  and tritium, two isotopes of hydrogen are brought together within the range of strong nuclear interactions to fuse a heavier element and free a neutron. According to Einstein's mass-energy equivalence theory, this process will generate substantial amounts of energy. In contrast to the nuclear fission, nuclear fusion reaction poses no risk of generating harmful radioactive substances. The by-products produced by the nuclear fusion are low-level and short-term, which can be easily disposed within a human life \cite{pironti2005fusion}. As such, it has become the best compromise between nature and the energy needs of human being.

\begin{figure}
\begin{center}
\includegraphics[width=3.5in]{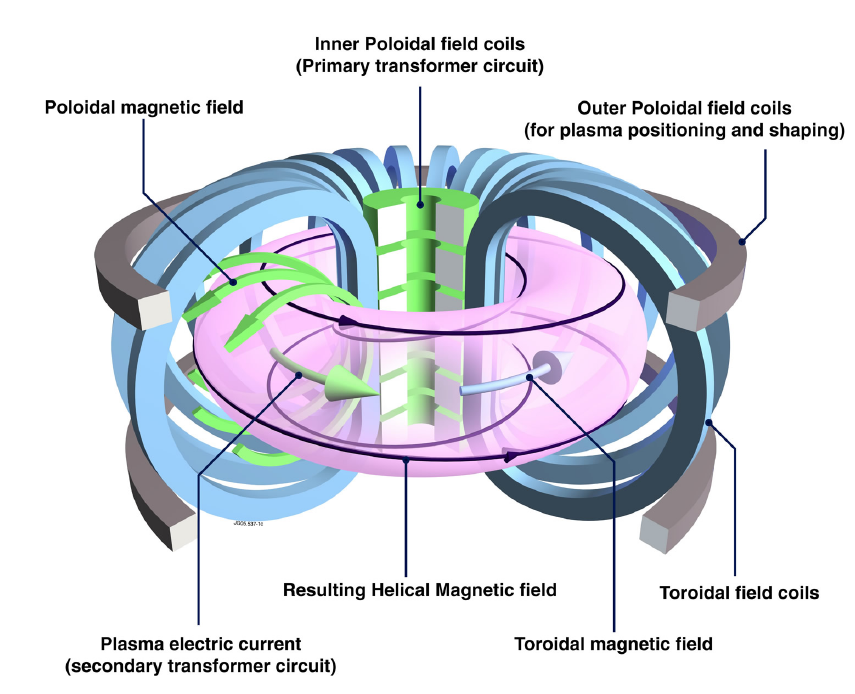}
\caption{Schematic of a tokamak chamber and magnetic profile (Source: EFDA-JET).}\label{tokamakschematic}
\end{center}
\end{figure}

Achieving a controlled fusion reaction on Earth is always a great challenge issue. There are two necessary conditions such as high temperature (about 100 million degrees) and high pressure must be satisfied to overcome the Coulomb barrier between the nuclei. Under these conditions of extremely high temperature and pressure, all the matter is in the plasma state,  which is a necessary part in the nuclear fusion. However, the hot plasma must be confined in order to prevent it from hitting the walls of the confining device. Fortunately, there is a magnetic confinement torus device called a tokamak (shown in Figure~\ref{tokamakschematic}), which can generate a helical magnetic field to confine the plasmas. The plasmas in tokamak can get squeezed by super-conducting magnets, thereby allowing fusion to occur.

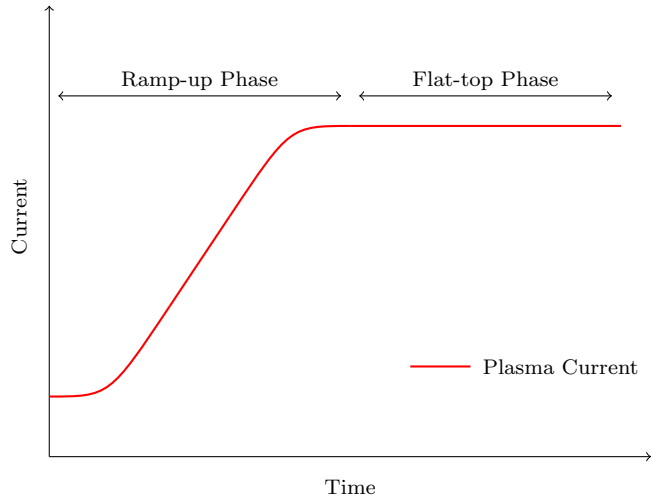
\begin{figure}
\begin{center}
\begin{tikzpicture}[scale=0.8]

\draw[->] (0,0) -- (10,0);
\draw[->] (0,0) -- (0,7.5);

\node[rotate=90] at (-0.5,4) {\footnotesize Current};
\node at (5,-0.5) {\footnotesize Time};

\draw[-,red,thick] (0,1) .. controls (1,1) .. (2,2.5);
\draw[-,red,thick] (2,2.5)--(3,4);
\draw[-,red,thick] (3,4) .. controls (4,5.5) .. (5,5.5);
\draw[-,red,thick] (5,5.5)--(9.5,5.5);

\draw[<->] (0.15,6)--(4.85,6);
\draw[<->] (5.15,6)--(9.35,6);
\node at (2.5,6.25) {\footnotesize Ramp-up Phase};
\node at (7.25,6.25) {\footnotesize Flat-top Phase};

\draw[-,thick,red] (6,1.5)--(7,1.5);
\node at (8.5,1.5) {\footnotesize Plasma Current};
\end{tikzpicture}
\caption{The plasma current evolution can be divided into two phases -- ramp-up phase and flat-top phase.} \label{PlasmaCurrent}
\end{center}
\end{figure}

In a tokamak, the attainment of a desired toroidal current profile is critical for high fusion gain and non-inductive sustainment of the plasma current for steady-state operation \cite{murakami2006progress,ou2007towards,witrant2007control}.
The evolution in time of the current profile in tokamak is related to the evolution of the poloidal magnetic flux, which is modeled by the magnetic diffusion equation, a parabolic partial differential equation (PDE) in the normalized cylindrical coordinate system \cite{hinton1976}.
For the plasma current evolution, it can be divided into two phases -- ramp-up phase and flat-top phase, as shown in Figure \ref{PlasmaCurrent}.
During the ramp-up phase, one possible approach to current profile control focuses in creating the desired current profile with the aim of maintaining this target profile during the subsequent phases of the discharge. This phase can be formulated as a finite-time optimal control problem for the magnetic flux diffusion PDE \cite{blum1989numerical}.
There has been attracting considerable attention in the literature to the problem of manipulating the current profile to achieve high performance.
In Moreau \cite{DMoreau}, the empirical models for current profile evolutions are derived using system identification techniques, then the models are used to synthesize a controller for the safety factor profile manipulation. In~\cite{ou2008design}, the extremum seeking approach~\cite{ariyur2003real} is used to compute the  optimal open-loop control trajectory during the ramp-up phase of the plasma discharge. This approach can handle quite complicated constraints without being trapped by local minima. A reduced-order model is obtained in~\cite{Xu} by using the proper orthogonal decomposition (POD) method \cite{xu2013low}, which can reduce the computational burden of the extremum seeking approach in~\cite{ou2008design} and make receding horizon control a feasible approach for online implementation~\cite{ou2011receding}. In \cite{chao2013}, the open-loop optimal control problem of the $q$-profile is solved in ramp-up tokamak plasmas using the minimal-surface theory. In \cite{Felici2012} and \cite{ren2015}, using the Galerkin method, a finite-dimensional ordinary differential equation (ODE) model based on the original magnetic diffusion PDE is obtained, then the optimization of open-loop actuator trajectories for the tokamak plasma profile control is solved by the nonlinear optimization algorithm.
There are also several advanced control and optimization approaches being investigated (e.g.,~\cite{boyer2013first,bribiesca2013strict,Dongen2014,shan2015start,witrant2007control,witvoet2012sawtooth}) due to advances in internal diagnostics and plasma actuation. However, many challenging research problems still remain open.

In the present work, using the framework proposed by Witrant \cite{witrant2007control} and Ou \cite{ou2007towards},
we consider the problem of obtaining a  predefined desired output current profile at the end of the current ramp-up phase.
We formulate the problem as a finite-time dynamic PDE-constrained optimization problem, in which the average density, total power, and plasma current are used as input control actuators to manipulate the profile shape. In this paper, we are aiming at proposing an effective framework for the solution of PDE-constrained optimization problem in tokamak plasmas.
We propose a numerical solution procedure directly to solve the original PDE-constrained optimization problem rather than discretizing the original PDE model over the space into a finite-dimensional system of ODEs (e.g., \cite{Dongen2014,Felici2012,ren2015,Xu}).
Applying the control parameterization approach, each control function is approximated by a linear combination of temporal basis functions with the constant coefficients to be determined by numerical optimization procedures such as sequential quadratic programming (SQP) based on the gradient optimization technique. A more user-friendly and efficient graphical user interface (GUI) is designed in MATLAB and the numerical simulation results are verified to demonstrate its applicability.

In fact, PDE-constrained optimization problem involves more applications, e.g.,  fluid pipeline \cite{Chen201554}, transport-reaction processes \cite{bendersky}, shape optimization \cite{haslinger2003}. There are usually two main approaches for solving PDE-constrained optimization problem: discretize-then-optimize framework or optimize-then-discretize framework. Designing the efficient solution of PDE-constrained optimization problem has a strong impact on these applications. In this paper, our proposed framework of combining existing PDEs and the numerical optimization solvers to solve PDE-constrained optimization problem has also the prospective to  target advanced control problems arising in more general spatial-temporal processes.

This paper is organized as follows. The formulation of the PDE-constrained optimization problem for the ramp-up current profile in tokamak plasmas is introduced in Section \ref{sec:cda}.
In Section~\ref{sec:CPM}, we give a numerical solution procedure to solve the PDE-constrained problem in which the control parameterization technique is used to approximate the control input variables by linear combinations of basis functions, thereby obtaining an approximate problem that can be solved using numerical optimization methods such as SQP. Then,  the details of framework for the optimization procedure are proposed. In Section~\ref{sec:Numerical}, numerical results for a plasma discharge in the DIII-D tokamak (under the shot \#129412) are presented. In Section~\ref{sec:conclusion}, we conclude the paper by summarizing our results and suggesting topics for further research.

\section{Problem Formulation}\label{sec:cda}
In a tokamak device, the plasma transport is governed by a set of PDEs that are nonlinearly coupled (with coupling strength depending on plasma scenarios).  Due to the assumptions made in \cite{witrant2007control} and \cite{ou2007towards}, the model used in this paper only contains  the magnetic flux PDE.
The dynamic behavior of the magnetic-flux profile $\psi(\hat{\rho},t)$ is described by the following parabolic PDE~\cite{Xu}:
\begin{equation}\label{model0}
\frac{1}{\vartheta_1(\hat{\rho})}\frac{\partial \psi(\hat{\rho},t)}{\partial t}=\frac{u_1(t)}{\hat{\rho}}\frac{\partial}{\partial \hat{\rho}}\left[\hat{\rho}D(\hat{\rho})\frac{\partial \psi(\hat{\rho},t)}{\partial \hat{\rho}}\right]+\vartheta_2(\hat{\rho})u_2(t),
\end{equation}
where $t$ is the time, $\hat{\rho} \in [0,1]$ is the normalized radius
and $\psi(\hat{\rho},t)$ is the poloidal magnetic flux around the tokamak. $\vartheta_1(\hat{\rho})$, $\vartheta_2(\hat{\rho})$ and $D(\hat{\rho}) $ are given functions that can be identified offline using experimental data.
The auxiliary functions $u_1(t)$, $u_2(t)$ depend on the physical actuators such as the average density $\bar{n}(t)$, the total power $P(t)$, the total plasma current $I(t)$. The nonlinear relationship between the auxiliary inputs $u_1(t)$, $u_2(t)$ and the actuators  $\bar{n}(t)$, $P(t)$, $I(t)$  satisfies the following equations:
\begin{subequations}\label{nip}
\begin{align}
&u_1(t)=\bar{n}(t)^{\frac{3}{2}}I(t)^{-\frac{3}{2}}P(t)^{-\frac{3}{4}},\\
&u_2(t)=I(t)^{-1}P(t)^{\frac{1}{2}}.
\end{align}
\end{subequations}
The starting values of input signals $\bar n(t)$, $I(t)$ and $P(t)$ are pre-specified before each plasma discharge
\begin{align}\label{NPI_bound}
\bar{n}(0)= \bar{n}_0, \quad
I(0)= I_0, \quad
P(0)= P_0,
\end{align}
where $\bar{n}_0$, $I_0$, $P_0$ are given values.
Furthermore, the open-loop input signals $\bar n(t)$, $I(t)$ and $P(t)$ also need to satisfy the following physical constraints:
\begin{align}\label{NPI_bound0}
\bar{n}_{\min}\leq \bar n(t)\leq \bar{n}_{\max}, \quad
I_{\min}\leq I(t)\leq I_{\max}, \quad
P_{\min}\leq P(t) \leq P_{\max},
\end{align}
and the terminal time actuator constraints:
\begin{align}\label{NPI_bound1}
\bar{n}(T)= \bar{n}_{\text{target}}, \quad
I(T)= I_{\text{target}}, \quad
P(T)= P_{\text{target}},
\end{align}
where $\bar{n}_{\min}$, $\bar{n}_{\max}$, $I_{\min}$, $I_{\max}$, $P_{\min}$, $P_{\max}$ and $\bar{n}_{\text{target}}$, $I_{\text{target}}$, $P_{\text{target}}$ are given constants. Any vector-valued function $\ \Theta=[\bar{n},I,P]^\top:[0,T] \rightarrow \mathbb{R}^3$ that satisfies the  constraints (\ref{NPI_bound0}) and (\ref{NPI_bound1}) is called an admissible control. Let $\Xi$ be the class of all such admissible controls.

The boundary conditions for the poloidal magnetic flux (\ref{model0}) are given by
\begin{equation}\label{boudary0}
\frac{\partial \psi(0,t)}{\partial \hat{\rho}}=0, \quad \frac{\partial \psi(1,t)}{\partial \hat{\rho}}=k I(t),
\end{equation}
where the $k$ is a given constant.
The initial condition for the magnetic
flux profile (\ref{model0}) is given by
\begin{equation} \label{intion0}
\psi(\hat{\rho},0)=\psi_0(\hat{\rho}).
\end{equation}
Furthermore, the system output is defined as the poloidal flux spatial derivative:
\begin{equation}\label{omege-eq}
\iota(\hat{\rho},t)=\frac{\partial\psi(\hat{\rho},t)}{\partial \hat{\rho}}.
\end{equation}
In the tokamak, our aim is to find the optimal open-loop input signals $\bar n(t)$, $I(t)$ and $P(t)$ so that the system output  $\iota(\hat{\rho},t)$  is as close as possible to a given target profile  $\iota_d(\hat\rho)$ at the terminal time $t=T$. Thus, we give the following cost functional to be minimized
\begin{equation}\label{costfunction-0}
\begin{aligned}
J(\hat{\rho})&=\frac{1}{N}\sum_{i=1}^N(\iota(\hat{\rho_i},T)-\iota_d(\hat{\rho_i}))^2+\varepsilon \sum_{i=1}^N \left(\frac{\partial\psi(\hat{\rho_i},T)}{\partial t}\right)^2,
\end{aligned}
\end{equation}
where $N$ denotes the number of discrete points in space within the interval $\hat{\rho}\in [0,1]$ for the normalized radius
and $\varepsilon$ is a given small positive constant. Note that in order to establish the appropriate current density profile for a quasi-static operation in the flat-top phase (as shown in Figure \ref{PlasmaCurrent}), we add the second term in (\ref{costfunction-0}) to penalize the terminal growth rate in the end of the ramp-up phase. By introducing the penalty term, we expect to maintain the achieved profile in the end of the ramp-up phase rather than just let it fly away in the beginning of the flat-top phase.
Now we state our dynamic optimization problem formally as follows.
\begin{problem0}
\textit{Given the poloidal magnetic flux (\ref{model0}) with boundary conditions (\ref{boudary0}) and initial condition (\ref{intion0}), find an admissible control $\Theta=[\bar{n}, I,P ]^\top \in \Xi$ such that the cost
functional (\ref{costfunction-0}) is minimized}.
\end{problem0}

\section{Numerical Solution Procedure}\label{sec:CPM}
\subsection{Control Parameterization}
In this section, we will use the control parameterization method \cite{linsurvey2013, Teo} to approximate the control inputs.
We first subdivide the whole time horizon $[0,T]$ into $p$ subintervals $[t_{k-1},t_k)$,  $k=1,2,\ldots,p$,
where $t_0=0$ and $t_p=T$.
For each subinterval $[t_{k-1},t_k)$,  $k=1,2,\ldots,p$, they satisfy the following constraints:
\begin{equation}\label{tc-defined}
\tau_{\min} \leq t_k-t_{k-1}\leq \tau_{\max}, \quad k=1,2,\dots,p.
\end{equation}
Here, $\tau_{\min}>0$ and $\tau_{\max}>0$ are the minimum and maximum subinterval durations, respectively.
Then, we use values $\bar n(t_k)$, $I(t_k)$ and $P(t_k)$ at each time knot points $t_k$, $k=1,2,\ldots,p$, to
parameterize the open-loop input signals $\bar n(t)$, $I(t)$ and $P(t)$. For each interior points $(t_{k-1},t_k)$,  $k=1,2,\ldots,p$,
we use the linear interpolations to approximate $\bar n(t)$, $I(t)$ and $P(t)$, i.e.,
\begin{subequations}\label{equ11}
\begin{align}
&\bar n_{linear}(t)\approx \bar n(t_{k-1})+\frac{\bar n(t_k)-\bar n(t_{k-1})}{t_k-t_{k-1}}(t-t_{k-1}),\\
& I_{linear}(t)\approx I(t_{k-1})+\frac{I(t_k)-I(t_{k-1})}{t_k-t_{k-1}}(t-t_{k-1}), \\
& P_{linear}(t) \approx P(t_{k-1})+\frac{P(t_k)-P(t_{k-1})}{t_k-t_{k-1}}(t-t_{k-1}).
\end{align}
\end{subequations}
Then, for the physical constraints (\ref{NPI_bound0}), we can easily obtain the following canonical bound constraints:
\begin{align}\label{NPI_bound-tk}
\bar{n}_{\min}\leq \bar n(t_k)\leq \bar{n}_{\max}, \quad
I_{\min}\leq I(t_k)\leq I_{\max}, \quad
P_{\min}\leq P(t_k) \leq P_{\max},
\end{align}
and the terminal time state constraints (\ref{NPI_bound1}) satisfy
\begin{align}\label{NPI_bound1-tk}
\bar{n}(t_p)= \bar{n}_{\text{target}}, \quad
I(t_p)= I_{\text{target}}, \quad
P(t_p)= P_{\text{target}}.
\end{align}
After we parameterize the open-loop input signals $\bar n(t)$, $I(t)$ and $P(t)$ as in (\ref{equ11}), the control input signals have been represented as
the linear functions of time $t$ and $\bar n(t_k)$, $I(t_k)$ and $P(t_k)$.
Define a parameterization vector
\begin{equation*}\label{opt_vector}
\Upsilon=(\bar n(t_1),n(t_2),\ldots,n(t_{p-1}),I(t_1),I(t_2),
\dots,I(t_{p-1}), P(t_1),P(t_2),\ldots,P(t_{p-1}))\in\Xi,
\end{equation*}
which contains all the parameters needed to be optimized. Now the Problem P is equivalent to the following problem.
\begin{problemN}
Given the poloidal magnetic flux (\ref{model0}) with boundary conditions (\ref{boudary0}) and initial condition (\ref{intion0}), find a control parameter vector $\Upsilon \in \Xi$ such that the cost
functional (\ref{costfunction-0}) is minimized subject to constraints (\ref{NPI_bound-tk}) and (\ref{NPI_bound1-tk}).
\end{problemN}
Problem P$_N$ is a canonical PDE-constrained optimization problem in which the parameters $\bar n(t_k)$, $I(t_k)$, $P(t_k)$, $i=1,2,\ldots,p-1$,
need to be selected optimally. Actually, we can solve such problems using the SQP method whose advantage is that the local extremal  solution can be obtained with a local quadratic convergence. To solve Problem P$_N$ using the SQP method, the key idea is to evaluate
the values of cost functional (\ref{costfunction-0}) and its gradients with respect to the decision parameters $\bar n(t_i)$, $I(t_i)$, $P(t_i)$, $i=1,2,\ldots,p-1$,
at different time instants, iteratively. Once we evaluate these simultaneously, then Problem P$_N$ can be solved using  the following gradient-based optimization algorithm.

\begin{algorithm}
\caption{Gradient-based optimization procedure for solving Problem P$_N$}
\SetKwData{Index}{Index}
\BlankLine
$\bf{Step~1}$. Set an initial guess starting vector $(\bar n^{(0)}(t_k), I^{(0)}(t_k), P^{(0)}(t_k))$.\\
\BlankLine
$\bf {Step~2}$. Solve the poloidal magnetic flux (\ref{model0}) with boundary conditions (\ref{boudary0}) \\~~~~~~~~~~~and initial condition (\ref{intion0}).\\
\BlankLine
$\bf {Step~3}$. Evaluate the value of cost functional (\ref{costfunction-0}) at the current state \\~~~~~~~~~~$(\bar n^{(m)}(t_k), I^{(m)}(t_k), P^{(m)}(t_k))$.\\
\BlankLine
$\bf {Step~4}$. Evaluate the gradients of cost functional (\ref{costfunction-0}) with respect to \\ ~~~~~~~~~~~parameters  $(\bar n^{(m)}(t_k), I^{(m)}(t_k), P^{(m)}(t_k))$,
using finite difference approximation.\\ ~~~~~~~~~~~For the bound constraints (\ref{NPI_bound-tk}) and actuator constraints (\ref{NPI_bound1-tk}), because they are\\~~~~~~~~~~~canonical forms in the nonlinear optimization problem, we can readily solve\\~~~~~~~~~~~them in the optimization program.\\
\BlankLine
$\bf {Step~5}$. Use the gradients information obtained in Step 4 to perform an optimality test.\\~~~~~~~~~~~If the current points is optimal, then stop; otherwise, go to Step 6.\\
\BlankLine
$\bf {Step~6}$. Use the gradients information obtained in Step 4 to calculate a search direction.
\BlankLine
$\bf {Step~7}$. Perform a line search to determine the optimal step length.\\
\BlankLine
$\bf {Step~8}$. Evaluate a new vector $(\bar n^{(m+1)}(t_k), I^{(m+1)}(t_k), P^{(m+1)}(t_k))$ and return to Step 2.
\label{algoritm1}
\end{algorithm}

The key steps in Algorithm \ref{algoritm1} are Steps 2-4. For the Steps 5-8, they can be performed automatically by existing nonlinear optimization solvers such as FMINCON combined with SQP method available in MATLAB. In the next section, we will focus on implementing the details of Algorithm \ref{algoritm1}.

\subsection{The Optimization Framework}\label{sec:frameopt}
\subsubsection{Cost Functional Implemention}\label{sec:costimplement}
The solver for the Problem P$_N$ is typically implemented in the following step.

Firstly, to evaluate the values of the cost functional (\ref{costfunction-0}) of Step 2 of Algorithm \ref{algoritm1}, we need to solve the poloidal magnetic flux (\ref{model0})
with boundary conditions (\ref{boudary0}) and initial condition (\ref{intion0}) (Step 2).
There are many numerical methods to solve the governing PDEs, i.e.,
finite element method (FEM), finite difference method (FDM). To implement Step 2 of Algorithm \ref{algoritm1}, we employ the PDEPE toolbox in MATLAB to compute the
value of magnetic-flux profile $\psi(\hat{\rho},t)$. The PDEPE is a toolbox for solving initial-boundary value problems for systems of
parabolic or elliptic PDEs in the one space variable and time variable. It can solve the following unified form of PDEs such as
\begin{equation}\label{pdepe-main}
\begin{aligned}
c\left(x,t,\phi(x,t),\frac{\partial \phi(x,t)}{\partial x}\right)\frac{\partial \phi(x,t)}{\partial t}&= x^{-m}\frac{\partial}{\partial x}\left(x^m f\left(x,t,\phi(x,t),\frac{\partial \phi(x,t)}{\partial x}\right)\right)\\
&\quad+s\left(x,t,\phi(x,t),\frac{\partial \phi(x,t)}{\partial x}\right),
\end{aligned}
\end{equation}
where $x \in [x_0,x_f]$ and $ t\in[t_0, t_f]$  denote the space and time variables, respectively, the parameter $m$ can be set 0, 1, or 2.
Furthermore, the boundary conditions using in PDEPE must be  the following unified form:
\begin{equation}\label{pdepe-bd}
p(x,t,\phi(x,t))+q(x,t)f\left(x,t,\phi(x,t),\frac{\partial \phi(x,t)}{\partial x}\right)=0,
\end{equation}
where $x=x_0$ or $x=x_f$.
Since the poloidal magnetic flux (\ref{model0}) and boundary conditions (\ref{boudary0}) are canonical forms expressed in the forms (\ref{pdepe-main}) and (\ref{pdepe-bd}), respectively, we can easily obtain
\begin{subequations}\label{pdepe-cfs}
\begin{align}
m&=1,\label{pdepe-m} \\
c\left(x,t,\phi(x,t),\frac{\partial \phi(x,t)}{\partial x}\right)&=\frac{1}{\vartheta_1(x)}, \\
f\left(x,t,\phi(x,t),\frac{\partial \phi(x,t)}{\partial x}\right)&=u_1(t)D(x)\frac{\partial \psi(x,t)}{\partial x},\\
s\left(x,t,\phi(x,t),\frac{\partial \phi(x,t)}{\partial x}\right)&=\vartheta_2(x)u_2(t).
\end{align}
\end{subequations}
Here, we use the variable $x$ denotes the variable $\hat{\rho}$ in (\ref{model0}).
On the left boundary condition $\hat{\rho}=0$, we have
\begin{equation}\label{pdepe-bdl}
\begin{aligned}
p_l(0,t,\phi)=0, \quad q_l(0,t)=\frac{1}{D(x)u_1(t)},
\end{aligned}
\end{equation}
and on the right boundary condition $\hat{\rho}=1$, we have
\begin{equation}\label{pdepe-bdr}
p_r(1,t,\phi)=-k I(t), \quad
q_r(1,t)=\frac{1}{D(x)u_1(t)}.
\end{equation}
In the designed program, we call the following function to compute the value $\psi(\hat{\rho},t)$,
\begin{lstlisting}[frame=single]
u = pdepe(m,@pde_fun,@pde_ic,@pde_bc,x_mesh,t_span,z);
\end{lstlisting}
where \texttt{u} denotes the returned value $\psi(\hat{\rho},t)$ computed by function \texttt{pdepe} and
\begin{itemize}
  \item parameter \texttt{m} corresponds to (\ref{pdepe-m}).
  \item subfunction \texttt{pde\_fun} corresponds to the terms $c\left(x,t,\phi(x,t),\frac{\partial \phi(x,t)}{\partial x}\right)$, $f\left(x,t,\phi(x,t),\frac{\partial \phi(x,t)}{\partial x}\right)$, and $s\left(x,t,\phi(x,t),\frac{\partial \phi(x,t)}{\partial x}\right)$ in (\ref{pdepe-cfs}), it has the following form:\\
      function [c, f, s] = \texttt{pde\_fun}(x, t, u, dudx, z).
  \item subfunction \texttt{ic\_fun} corresponds to the initial condition (\ref{intion0}), which uses the following form:\\
      function \texttt{u0} = \texttt{pde\_ic(x)}.
  \item subfunction \texttt{bc\_fun} corresponds to the terms $p(x,t,\phi(x,t))$, $q(x,t)$ in (\ref{pdepe-bdl}) and (\ref{pdepe-bdr}), it uses the following form:\\
      function [pl, ql, pr, qr] = \texttt{pde\_bc}(xl, ul, xr, ur, t, z).
  \item \texttt{x\_mesh} denotes a vector $[x_0, x_1,\ldots, x_f]$ satisfying the condition $x_0  < x_1 < \ldots<x_f$.
  \item \texttt{t\_span} denotes a vector $[t_0, t_1,\ldots, t_f]$ satisfying the condition $t_0 < t_1 < \ldots < t_f$.
  \item \texttt{z} denotes the to-be-optimized vector $(\bar n(t_1),\bar n(t_2),\ldots,\bar n(t_{p-1}),I(t_1),I(t_2),
\dots,I(t_{p-1}), P(t_1),P(t_2),\\ \ldots,P(t_{p-1}))$, it will be transferred to the subfunctions \texttt{ic\_fun} and \texttt{bc\_fun}.
\end{itemize}
Once we set all the parameters and functions in function \texttt{pdepe}, it will return the solution of $\psi(\hat{\rho}_i,t_i)=u(x_i,t_i)$ at the space nodes $x_1,x_2,\ldots,x_f$ and time nodes $t_1,t_2,\ldots,t_f$. Note that the cost functional (\ref{costfunction-0}) contains the terms $\iota(\hat{\rho_i},T)$ and $\frac{\partial\psi(\hat{\rho}_i,T)}{\partial t}$. We use the forward finite difference method to approximate $\iota(\hat{\rho_i},T)$ in the domain $\hat{\rho}_i\in (0,1)$:
\begin{equation}\label{equ:finite1}
\iota(\hat{\rho_i},T)=\frac{\partial\psi(\hat{\rho_i},T)}{\partial \hat{\rho}}\approx\frac{\psi(\hat{\rho}_i+h,T)-\psi(\hat{\rho}_i,T)}{h},
\end{equation}
where $h$ denotes the space size. On the left boundary $\hat{\rho}=0$, using the Taylor series expansion, we can easily obtain
\begin{equation}\label{equ:finite2}
\psi(0+h,T)\approx \psi(0,T)+h\frac{\partial\psi(0,T)}{\partial \hat{\rho}}+\frac{h^2}{2!}\psi''(0,T),
\end{equation}
where $\psi''(0,T)$ denotes the second derivative at $\hat{\rho}=0$ and can be obtained using the forward finite difference method
\begin{equation}\label{equ:finite3}
\psi''(0,T)=\frac{\psi(0+2h,T)-2\psi(0+h,T)+\psi(0,T)}{h^2}.
\end{equation}
Substituting (\ref{equ:finite3}) into (\ref{equ:finite2}), we obtain
\begin{equation}\label{equ:finite4}
\iota(0,T)=\frac{\partial\psi(0,T)}{\partial \hat{\rho}}\approx\frac{4\psi(0+h,T)-2\psi(0+2h,T)-3\psi(0,T)}{2h}.
\end{equation}
Similarly, we can obtain the following approximation
\begin{equation}\label{equ:finite5}
\iota(1,T)=\frac{\partial\psi(1,T)}{\partial \hat{\rho}}\approx\frac{3\psi(1,T)-4\psi(1-h,T)+\psi(1-2h,T)}{2h},
\end{equation}
and
\begin{equation}\label{equ:finite6}
\frac{\partial\psi(\hat{\rho}_i,T)}{\partial t}\approx\frac{3\psi(\hat{\rho}_i,T)-4\psi(\hat{\rho}_i,T-\tau)+\psi(\hat{\rho}_i,T-2\tau)}{2\tau},
\quad \hat{\rho}_i \in [0,1].
\end{equation}
where $\tau$ denotes the time step size.
By using (\ref{equ:finite1}), (\ref{equ:finite4}), (\ref{equ:finite5}) and (\ref{equ:finite6}), then the cost functional (\ref{costfunction-0}) can be evaluated sequentially by the input parameter vector $\Upsilon$.

\subsubsection{Parameters Optimization Implemention}\label{sec:optimalimp}
In section {\ref{sec:costimplement}}, we have implemented the framework of evaluating the cost functional (\ref{costfunction-0}). In this section, we will mainly focus on implementing the framework of optimizing the parameter vector $\Upsilon$ by using the existing nonlinear optimization solvers such as FMINCON combined with SQP algorithm in MATLAB.

The SQP algorithm solves a set of constrained quadratic optimization sub-problems obtained by quadratically approximating the cost function and linearizing the constraints. For the SQP algorithm, it requires the gradients of both the cost function and constraint functions with respect to the to-be-optimized parameters at the current states. 
For Problem P$_N$, it is possible to use the finite difference method to calculate these gradients. Fortunately, evaluating these gradients by finite difference can be done automatically using the optimization function FMINCON in MATLAB. The FMINCON has combined with SQP algorithm, which can be called as necessary by setting options in the FMINCON function.
The FMINCON attempts to find a constrained minimum of a cost function starting at an initial guess point. It can solve the nonlinear optimization problem specified by the following form:
\begin{subequations}\label{sqp-eqution}
\begin{align}
\min_p \quad &J(\dot{x}(t_f),x(t_f),p),\\
\text{subject to:} & \nonumber \\
 f(\dot{x}(t),x(t),p)&=0,\\
 c(x(t_f))&\leq 0,\quad c_{eq}(x(t_f))=0,\\
 A \cdot p &\leq b,\quad A_{eq} \cdot p = b_{eq},\\
 lb\leq  & p   \leq ub,
\end{align}
\end{subequations}
where $p$ is the to-be-optimized parameter vector and $x(t)$ denotes the system state, $x$, $b$, $b_{eq}$, $lb$, and $ub$ are vectors, $A$ and $A_{eq}$ are matrices, $c(x(t_f))$ and $c_{eq}(x(t_f))$ are functions that return vectors, and $J(\dot{x}(t_f),x(t_f),p)$ is a cost function that returns a scalar. Note that $f(\dot{x}(t),x(t),p)$, $c(x(t_f))$, and $c_{eq}(x(t_f))$ can be nonlinear functions.

To optimize the parameter vector $\Upsilon$ in our program, we call the function FMINCON as the following form:
\begin{lstlisting}[frame=single]
[dec_var,obj_val] = fmincon(@objfun(z), z0, A, b, Aeq, beq, lb, ub, @nonlcon, options);
options = optimset(`GradObj', `off', `Algorithm', `sqp');
\end{lstlisting}
The above function will return the cost functional (\ref{costfunction-0}) value \texttt{obj\_val} at the optimal solution \texttt{dec\_var}. The subfunction and parameters in the function fmincon are set
as:
\begin{itemize}
  \item function \texttt{objfun(z)} denotes the objective function (\ref{costfunction-0}) evaluated at parameter vector \texttt{z}. Since the cost functional value (\ref{costfunction-0}) in Section \ref{sec:costimplement} has been evaluated successfully using the function \texttt{pdepe}, we can specify it into the function \texttt{objfun(z)}.
  \item \texttt{z0} denotes the initial guess vector for \texttt{z}.
  \item \texttt{A}, \texttt{b}, \texttt{Aeq}, and \texttt{beq} denote the linear constraint matrices $A$ and $A_{eq}$, and their corresponding vectors $b$ and $b_{eq}$ in (\ref{sqp-eqution}). Since there are no linear constraints in Problem P$_N$, we set them into null in our program.
  \item \texttt{lb} and \texttt{ub} denote the vector of lower and upper bounds for parameter vector \texttt{z}, respectively. Here, we can set them using the constraints (\ref{NPI_bound-tk}).
  \item \texttt{nonlcon} denotes the nonlinear constraint function, since there are no nonlinear constraints in Problem P$_N$, we set it into null in the program.
  \item ``\texttt{options}'' denotes an optimization options structure. Here, we use the SQP algorithm to optimize the Problem P$_N$. Since the cost functional gradients can be evaluated by the finite difference method, we set the command ``\texttt{GradObj}'' is ``\texttt{off}''. This opinion will cause the function \texttt{fmincon} to estimate the  gradients of the cost function using finite differences method.
\end{itemize}

\begin{figure}
\begin{center}
\includegraphics[width=4.5in]{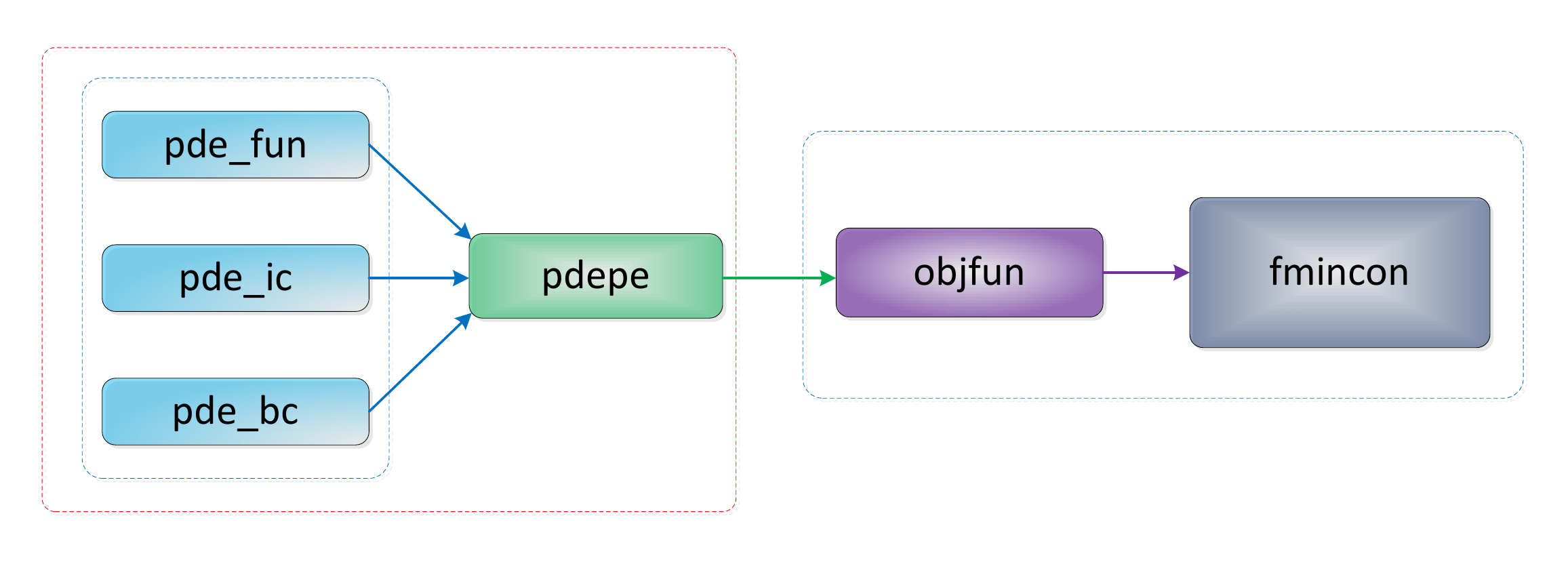}
\caption{Functions input.}\label{fig:funsinputs}
\end{center}
\end{figure}

After setting all the functions and parameters in functions \texttt{fmincon} and \texttt{pdepe}, Problem P$_N$ will be solved successfully. The relationship of functions input is shown in Figure \ref{fig:funsinputs}.

\begin{remark}
In the present work, the finite difference method is used to evaluate the gradients of the cost function. However, there are also other methods such as  adjoint method or sensitivity method to evaluate the cost functional gradients. For example, in our previous work \cite{ren2015}, Galerkin method is used to obtain a finite-dimensional ODE model based on the original magnetic diffusion PDE, then the gradients of the cost function with respect to the decision parameters can be evaluated analytically based on the forward sensitivity analysis. The computation of the cost functional gradient using these alternative approaches such as the adjoint or sensitivity methods can be readily incorporated into current framework. This will be our next step to integrate our previous work on different gradient computation modules \cite{ren2015} to unify the numerical dynamic optimization process.
\end{remark}

\subsubsection{GUI Implemention}
\begin{figure}
\begin{center}
\includegraphics[width=4.0in]{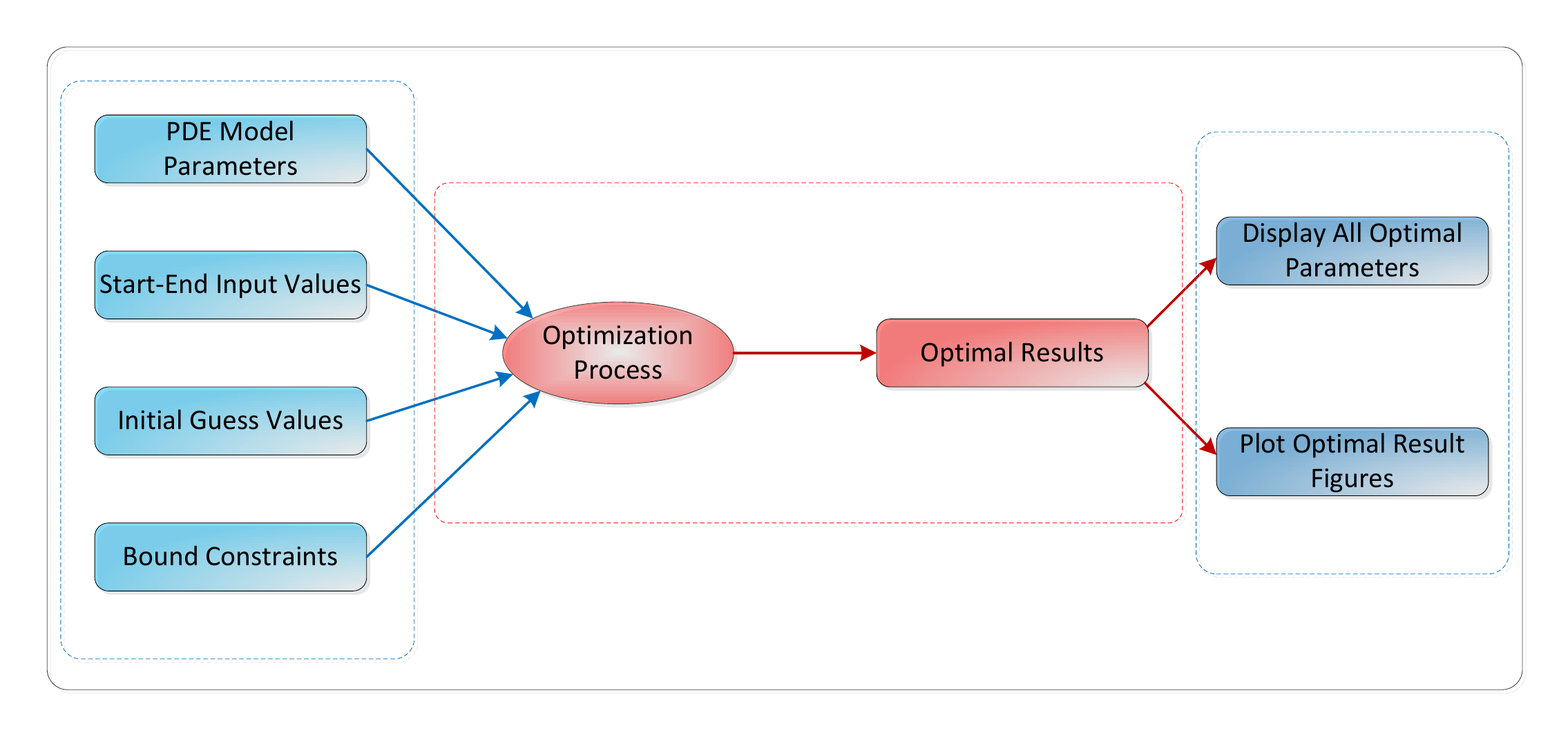}
\caption{System parameters inputs and optimization processing.}\label{fig:guiframe}
\end{center}
\end{figure}

\begin{figure}
\begin{center}
\includegraphics[width=3.5in]{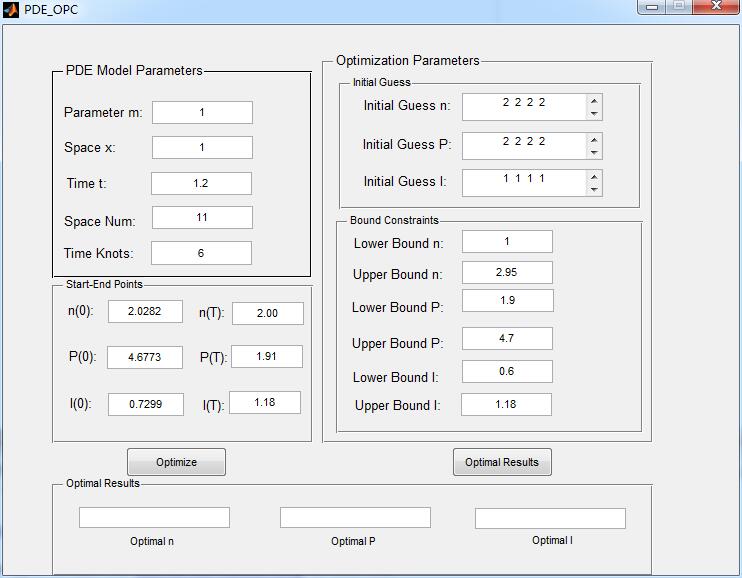}
\caption{GUI for system parameters inputs and output results (Source code: http://pan.baidu.com/s/1qWLyCDu).}\label{fig:gui}
\end{center}
\end{figure}

Based on Sections \ref{sec:costimplement}-\ref{sec:optimalimp}, there are four types of  parameters in the PDE system and optimization processing need to be required. We give an illustration of these parameters in Figure \ref{fig:guiframe}.
\begin{itemize}
  \item \texttt{PDE Model Parameters} include the value of parameter $m$, space variable $\hat{\rho}$, terminal time $T$, the number of discrete points in space $N$, and the number of time knots $k$ during the whole time horizon $[0,T]$.
  \item \texttt{Start-End Input Values} define the start values $\bar n(0)$, $I(0)$, $P(0)$ and the terminal conditions $\bar n(T)$, $I(T)$ and $P(T)$.
  \item \texttt{Initial Guess Values} define the initial guess starting vector $(\bar n^{(0)}(t_k), I^{(0)}(t_k), P^{(0)}(t_k))$ for each knot.
  \item \texttt{Bound Constraints} define the lower and upper bound constrains $\bar{n}_{\min}$, $\bar{n}_{\max}$, $P_{\min}$, $P_{\max}$, $I_{\min}$, $I_{\max}$ for the control parameters $\bar n(t_k)$, $P(t_k)$ and $I(t_k)$.
\end{itemize}
Once these parameters are set availably in the program, the \texttt{Optimization Process} module will check whether or not these parameters are within the allowable ranges, then the \texttt{Optimization Process} module will compute the values of cost function and their corresponding gradients, and then call the nonlinear optimization solver FMINCON to perform the optimization process and output the optimal results. The \texttt{Optimal Results} module will display all the optimal results, which include the optimal $\bar n(t)$, $I(t)$, $P(t)$ and $\iota(\hat{\rho},T)$, as well as plotting all these results. We implement the GUI for all these necessary data set in the program, as shown in Figure \ref{fig:gui}. The Optimize button will execute the \texttt{Optimization Process} module in Figure \ref{fig:guiframe} and the \texttt{Optimal Results} button will display all the optimal inputs $\bar n(t_k)$, $P(t_k)$ and $I(t_k)$ at each time knot points $t_k$, $k=1,2,\ldots,p,$ in the edit text boxes embedded in the GUI as well as displaying all the figures of the optimal results.

\section{Numerical Simulations} \label{sec:Numerical}
\begin{figure*}
\begin{center}
\begin{minipage}[c]{0.5\textwidth}
\centering\includegraphics[width=1\columnwidth]{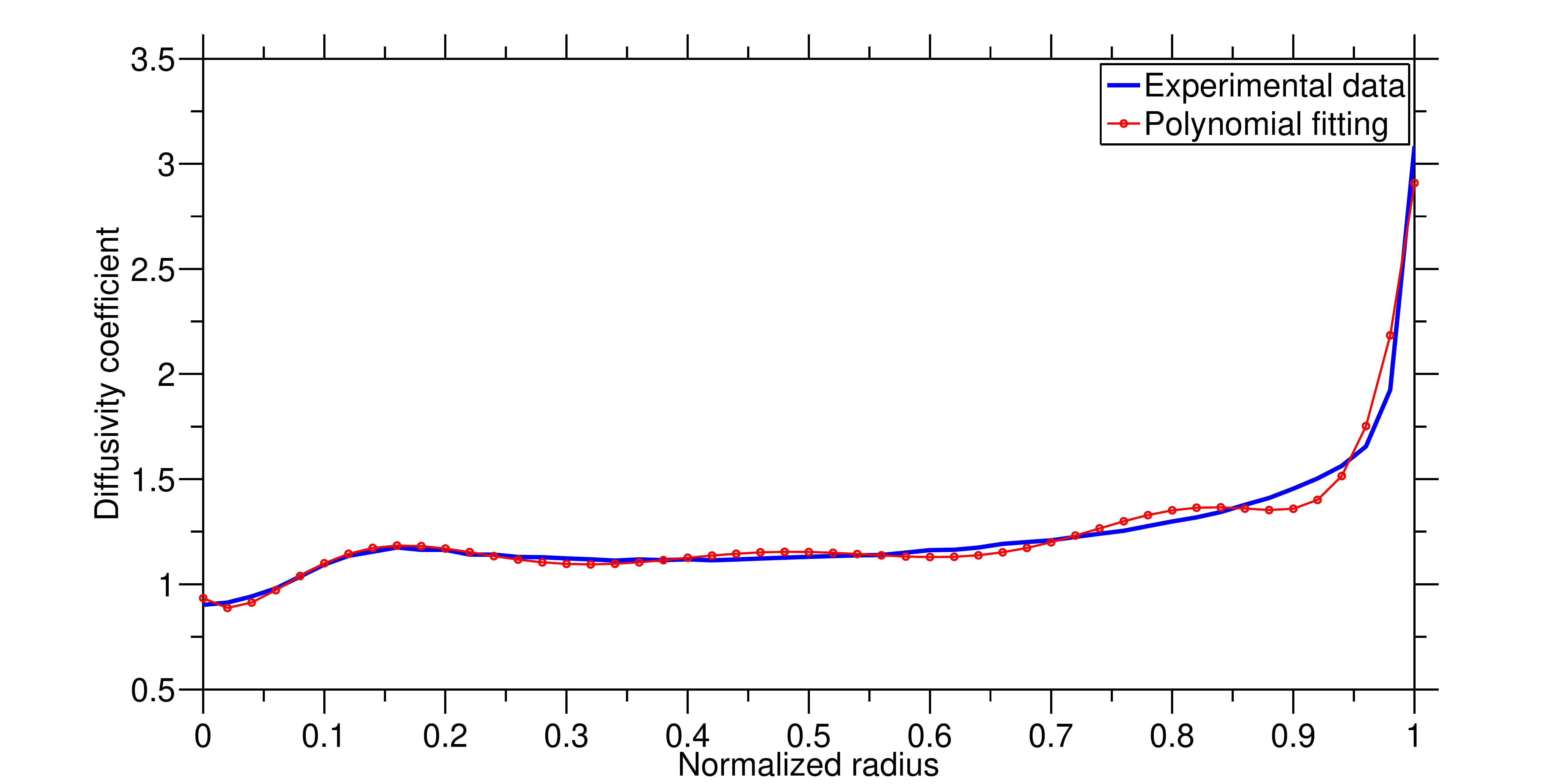}%
\renewcommand{\figurename}{Figure}
\caption{Diffusivity coefficient function $D(\hat{\rho})$. }
\label{fig:Diffusivity}
\end{minipage}%
\begin{minipage}[c]{0.5\textwidth}
\centering\includegraphics[width=1\columnwidth]{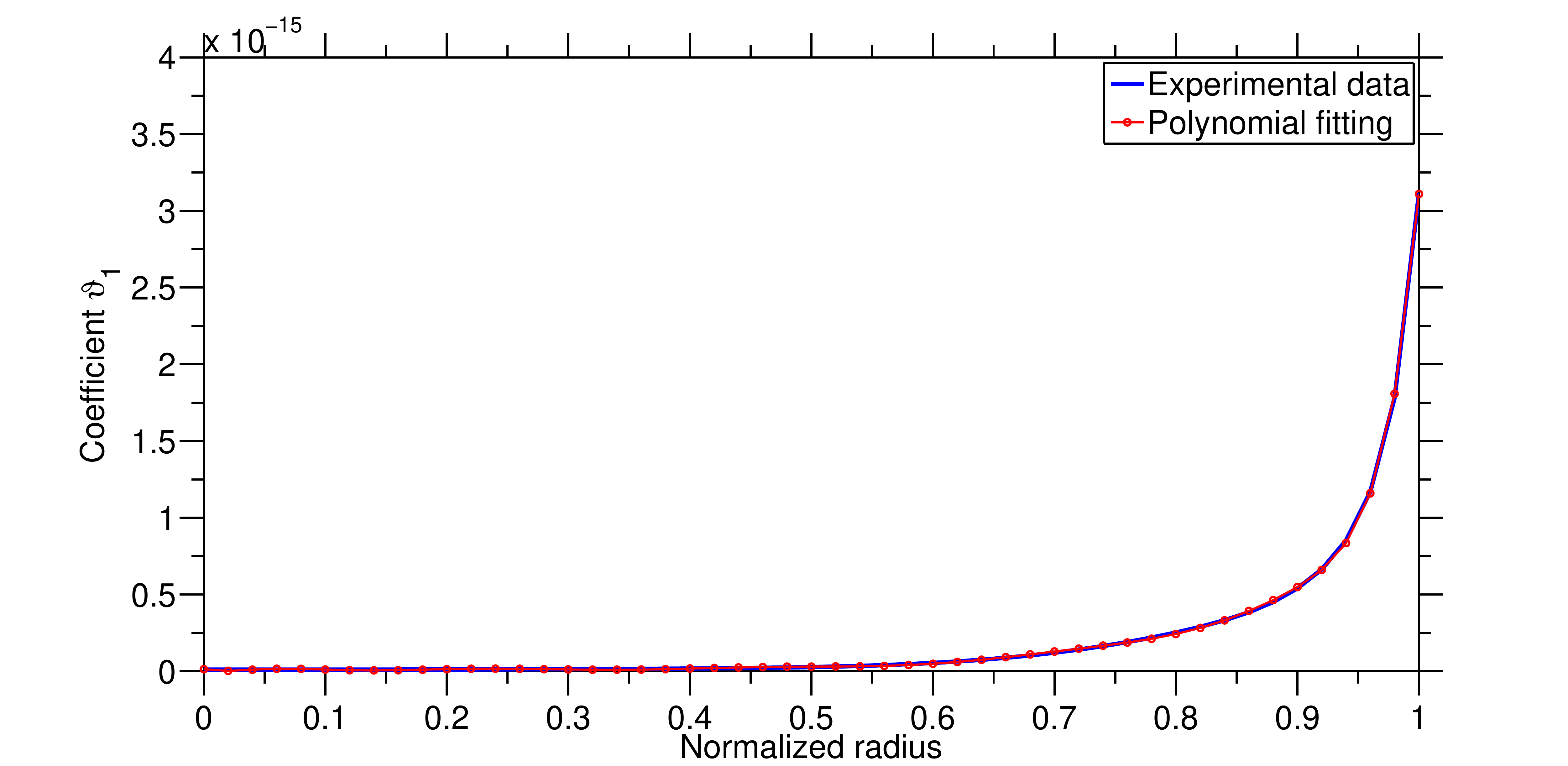}%
\renewcommand{\figurename}{Figure}
\caption{Coefficient function $\vartheta_1(\hat{\rho})$. }
\label{fig:vartheta1}
\end{minipage}
\end{center}
\end{figure*}

\begin{figure*}
\begin{center}
\begin{minipage}[c]{0.5\textwidth}
\centering\includegraphics[width=1\columnwidth]{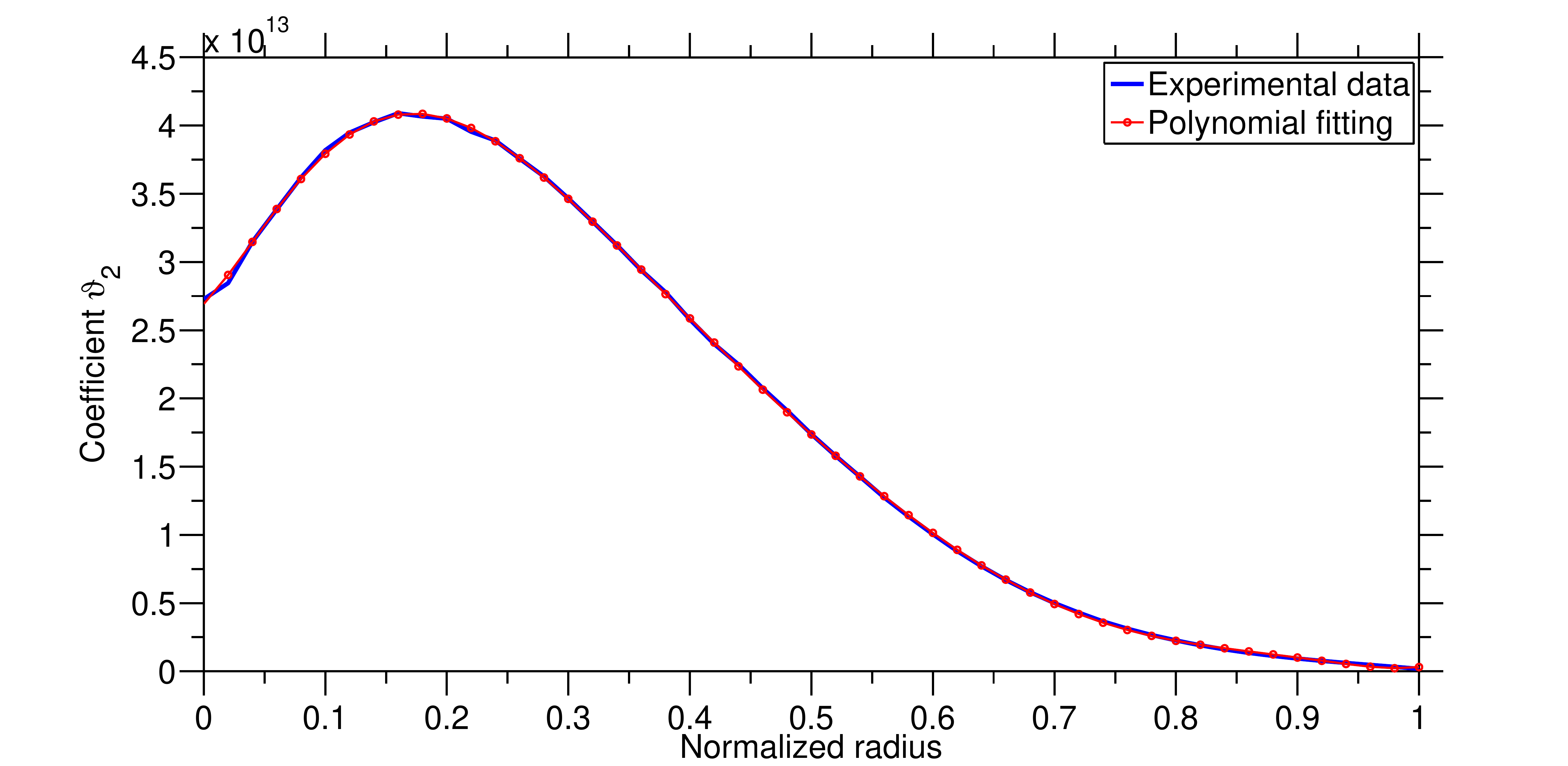}
\renewcommand{\figurename}{Figure}
\caption{Coefficient function $\vartheta_2(\hat{\rho})$. }
\label{fig:vartheta2}
\end{minipage}%
\begin{minipage}[c]{0.5\textwidth}
\centering\includegraphics[width=1\columnwidth]{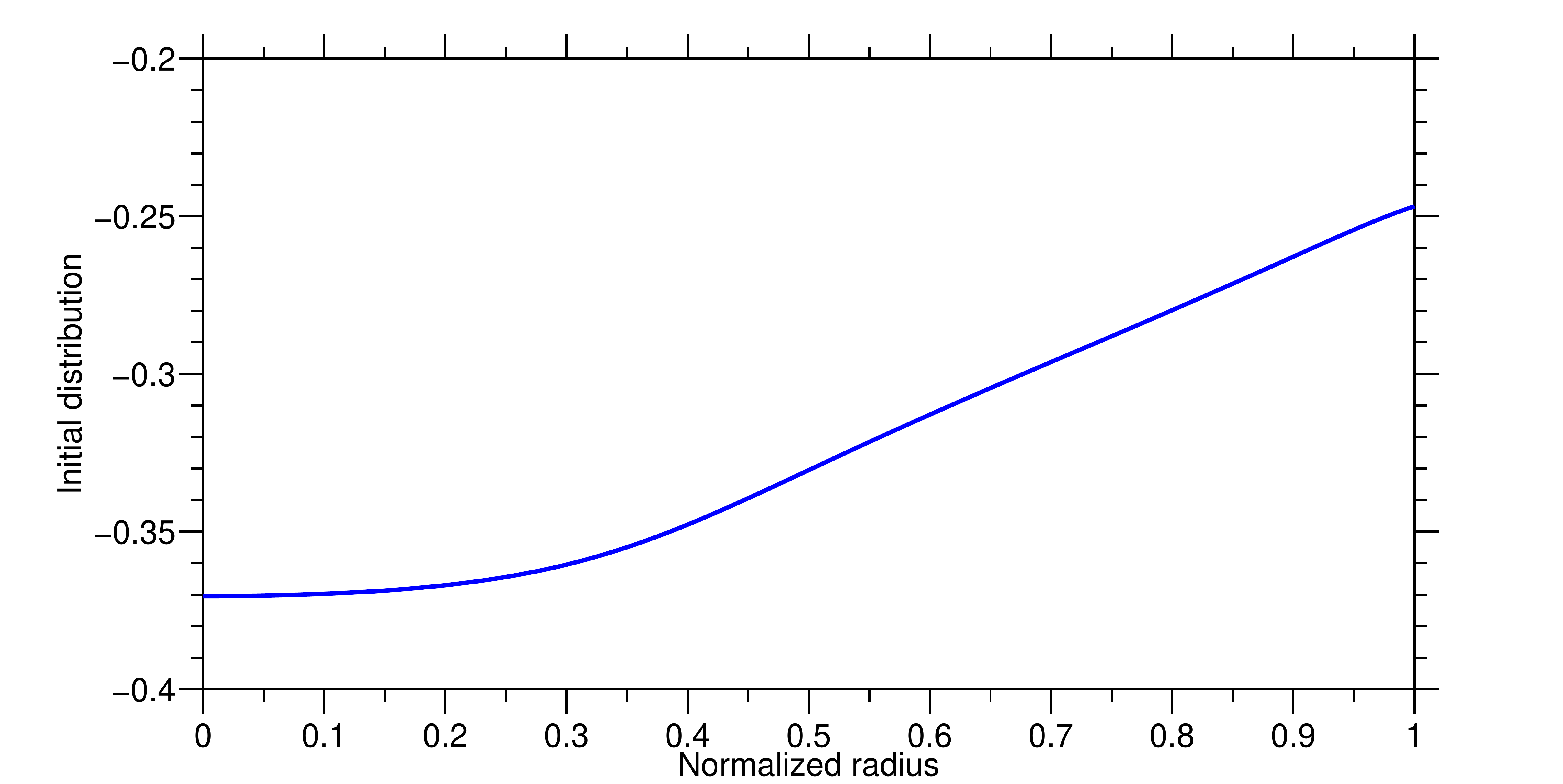}
\renewcommand{\figurename}{Figure}
\caption{The initial distribution $\psi_0(\hat{\rho})$. }
\label{fig:initialcond}
\end{minipage}
\end{center}
\end{figure*}
We now apply the computational method proposed in Sections 3 to an example. This example, which comes from \cite{Xu}, is based on experimental data from the DIII-D tokamak in San Diego, California. The functions $D(\hat{\rho})$, $\vartheta_1(\hat{\rho})$, and $\vartheta_2(\hat{\rho})$ in the PDE model (\ref{model0}) are shown in Figures \ref{fig:Diffusivity}-\ref{fig:vartheta2}. The initial magnetic flux profile is taken from shot $\#129412$ from the DIII-D tokamak (see Figure \ref{fig:initialcond}). The constant parameter $k$ in boundary condition (\ref{boudary0}) is given $1.0996 \times 10^{-7}$, which also comes from the experimental data.
For the input bounds and initial conditions,  they are given in Table \ref{tab:alpha10}.
\begin{table}
\centering
\caption{Input bounds and initial conditions and terminal constraints.}\label{tab:alpha10}
\begin{tabular}{ll}
  \toprule
  Variable & Value \\
  \toprule
  $\bar{n}_0$ & $2.03~[10^{19}\mathrm{m^{-3}}]$ \\
  $\bar{n}_{\min}$ & $1.00~[10^{19}\mathrm{m^{-3}}]$ \\
  $\bar{n}_{\max}$ & $2.95~[10^{19}\mathrm{m^{-3}}]$ \\
  $\bar{n}_{\text{target}}$ & $2.00~[10^{19}\mathrm{m^{-3}}]$ \\
  $I_0$ & $0.73~\mathrm{[MA]}$ \\
  $I_{\min}$  & $0.6~\mathrm{[MA]}$ \\
  $I_{\max}$  & $1.18~\mathrm{[MA]}$ \\
  $I_{\text{target}}$ & $1.18~\mathrm{[MA]}$ \\
  $P_0$ & $4.67~\mathrm{[MW]}$ \\
  $P_{\min}$  & $1.90~\mathrm{[MW]}$ \\
  $P_{\max}$  & $4.70~\mathrm{[MW]}$ \\
  $P_{\text{target}}$ & $1.91~\mathrm{[MW]}$ \\
  \hline
\end{tabular}
\end{table}

\begin{figure}
\begin{center}
\includegraphics[width=5.0in]{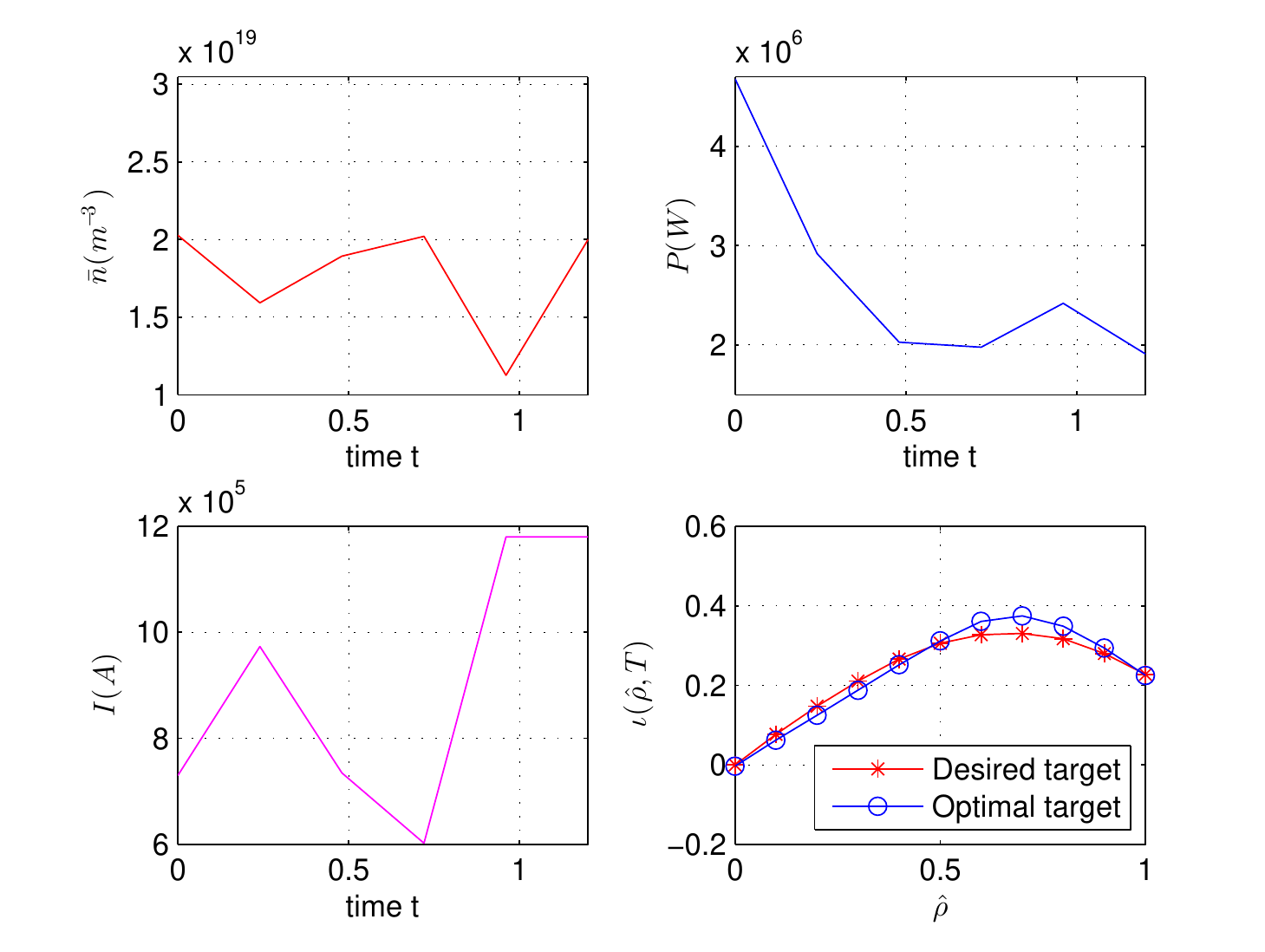}
\caption{Optimal controls for target profile 1 ($N=6$).}\label{fig:targe16}
\end{center}
\end{figure}

\begin{figure}
\begin{center}
\includegraphics[width=5.0in]{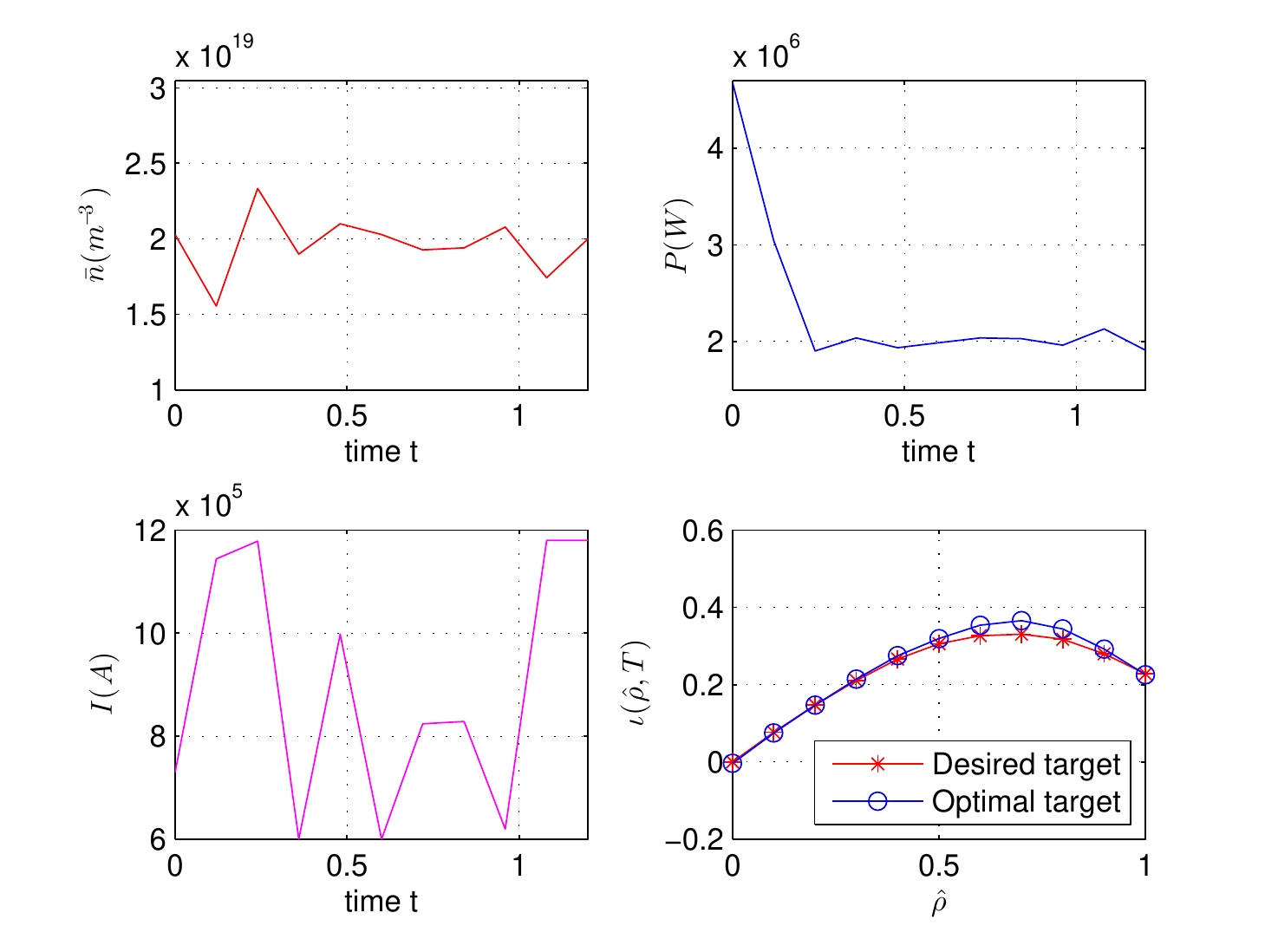}
\caption{Optimal controls for target profile 1 ($N=11$).}\label{fig:targe111}
\end{center}
\end{figure}

\begin{figure}
\begin{center}
\includegraphics[width=5.0in]{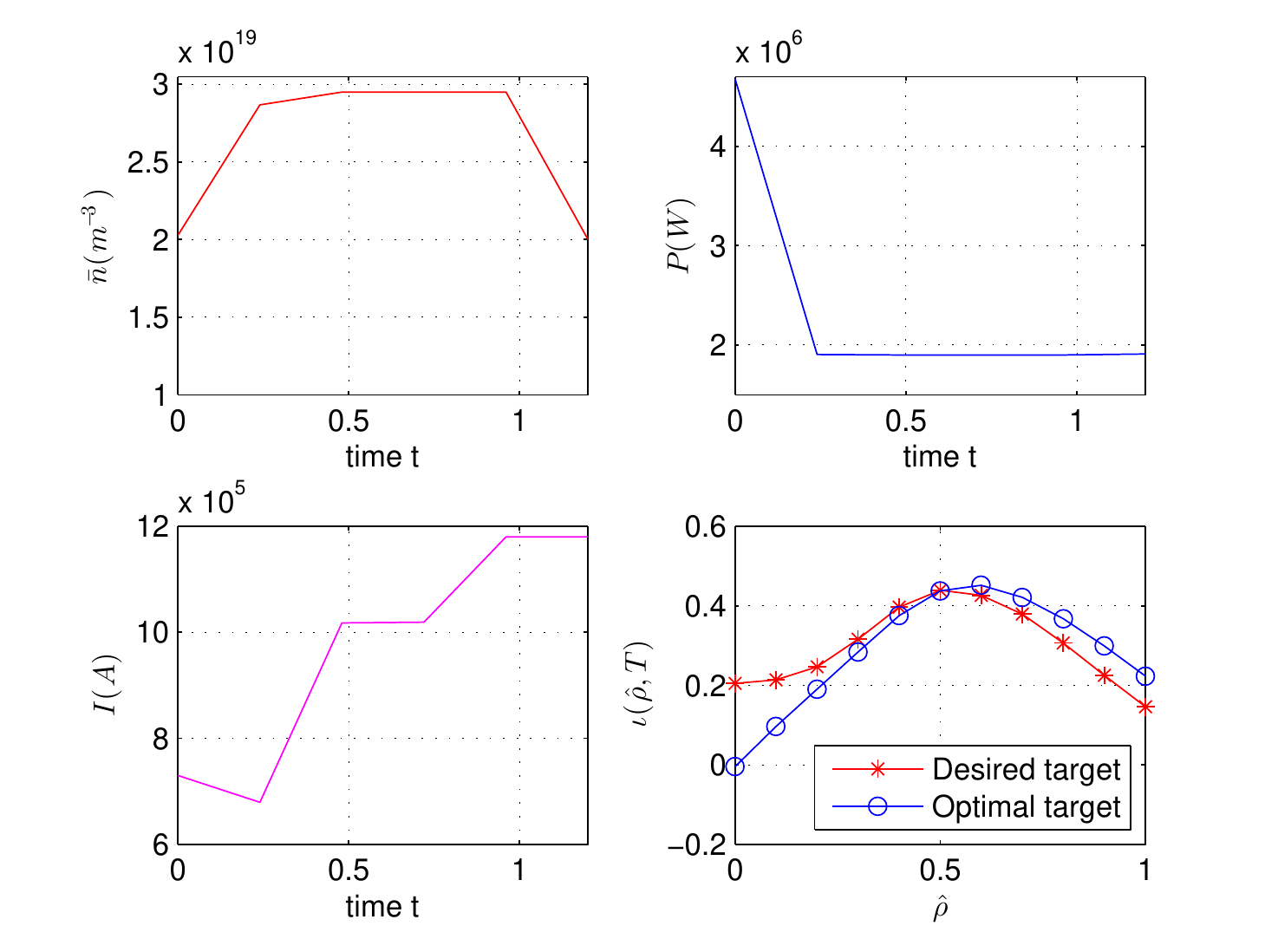}
\caption{Optimal controls for target profile 2 ($N=6$).}\label{fig:targe26}
\end{center}
\end{figure}

\begin{figure}
\begin{center}
\includegraphics[width=5.0in]{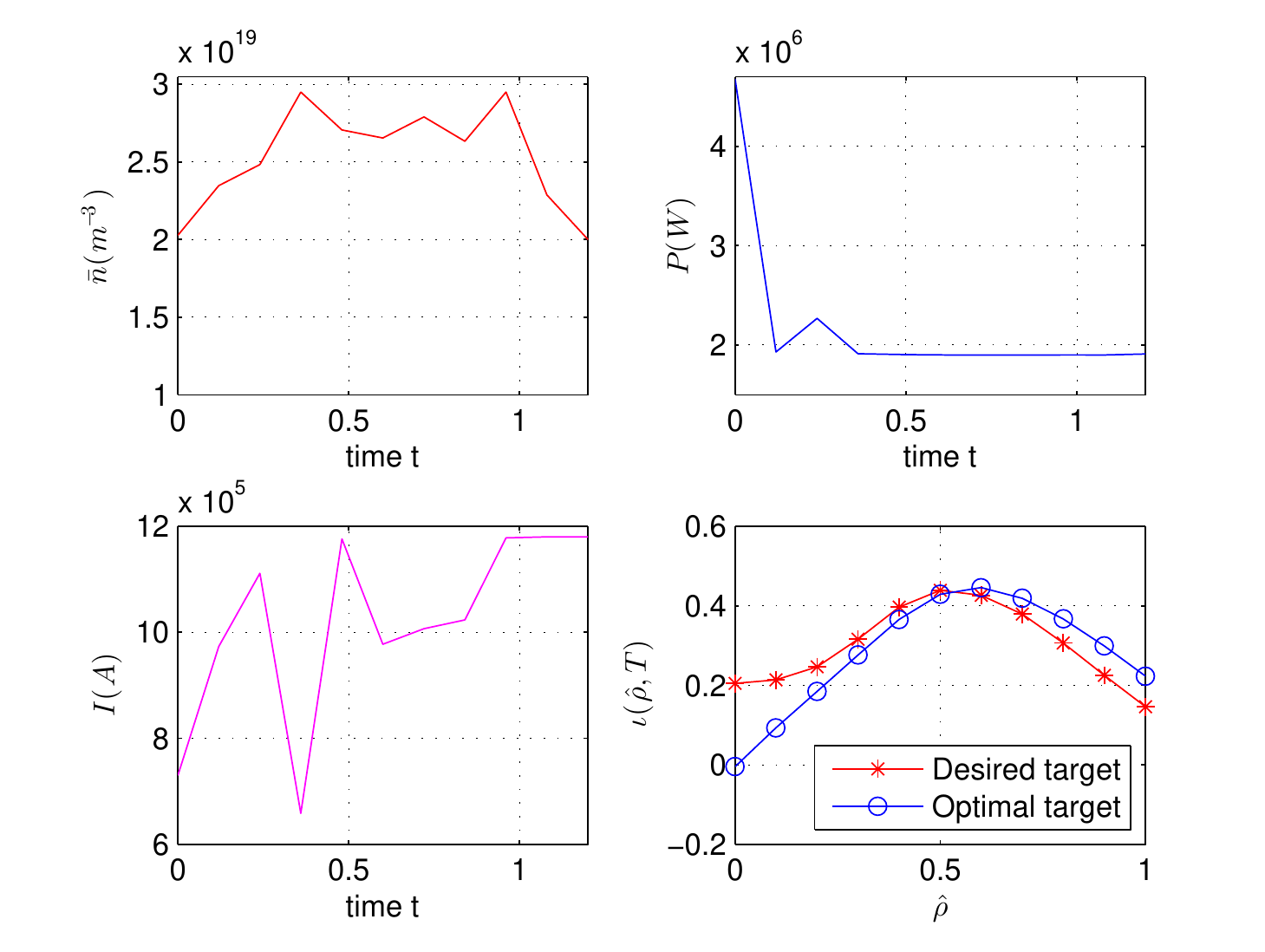}
\caption{Optimal controls for target profile 2 ($N=11$).}\label{fig:targe211}
\end{center}
\end{figure}
Our numerical simulation study was carried out within the MATLAB programming environment running on a personal computer with the following configuration: Intel Core i7-2600 3.40GHz CPU, 4.00GB RAM, 64-bit Windows 7 Operating System.  The MATLAB code implements the gradient-based optimization procedure described in Section 3 by combining FMINCON. We considered two target profiles $\iota_d(\hat{\rho})$ in our simulations: the first target profile is generated using the  experimental input data in \cite{Xu}; the second target profile  is generated using the experimental output flux data in \cite{ou2008design} (see Figure 20(a) in \cite{ou2008design}). In the following two numerical simulation examples, the value of penalty term parameter $\varepsilon$ in cost functional (\ref{costfunction-0}) is chosen $\varepsilon=0.05$. Note that this chosen value is enough to penalize the terminal growth rate in the end of the ramp-up phase so as to establish the appropriate current density profile for a quasi-static operation. The penalty parameter can also be regulated manually during the processes of numerical experiment but need to consider the efficiency of optimization.

\subsection{Example 1: Optimal Control for Target Profile 1}
For the target profile 1 in the example, we apply the control parameterization technique proposed in Section \ref{sec:CPM} to subdivide the time interval $[0,T]=[0,1.2\text{s}]$ into $p$
intervals. The control input functions $\bar n(t)$, $I(t)$, $P(t)$ are approximated by the piecewise-linear functions with break-points at $t_1, t_2, \ldots, t_{p-1}$, where $t_0=0$ and $t_p=1.2\text{s}$. We first considered $p=6$ so that the equidistant switching times are $\tau_{\min}=\tau_{\max}=0.2$. The optimal controls and optimal flux profiles generated by our proposed method are shown in Figure \ref{fig:targe16}. The results show that our proposed numerical optimization method can drive the final $\iota$-profile to within close proximity of the desired target profile. We also increase $p=11$ and the optimal results are shown in Figure \ref{fig:targe111}. Note that increasing $p$ results in reduced matching error, as expected.

\subsection{Example 2: Optimal Control for Target Profile 2}
For the target profile 2 in the example, we also considered $p=6,11$ to subdivide the time interval $[0,1.2\text{s}]$. The optimal time evolutions for the three input signals $\bar n(t)$, $I(t)$, $P(t)$ and the corresponding output trajectory $\iota(\hat{\rho},T)$ are shown in Figures \ref{fig:targe26}-\ref{fig:targe211}. The results also validate that our proposed numerical optimization method are effective.

\begin{remark}
For the numerical simulations, note that the SQP algorithm (or any other nonlinear programming algorithm) is designed to find local optima.  We can never guarantee that a local optimum as found by SQP or other algorithms is also one global optimum. The risk of finding the local minima can be reduced efficiently by choosing different initial conditions. During the numerical simulation processes, we also try different initial guess conditions and choose the satisfied results that the optimization procedure yields an improvement over the initial conditions.
\end{remark}
\section{Conclusion and Future Research} \label{sec:conclusion}

In this paper, a simplified dynamic model describing the evolution of the poloidal flux has been used. A finite-time PDE-constrained optimal control problem is proposed  arising during the ramp-up phase of a tokamak plasma and solved successfully under our proposed numerical optimization framework. We designed a user-friendly GUI embedded in FMINCON and SQP algorithm  to carry out the computations for solving the PDE-constrained optimal problem directly (Source Code: http://pan.baidu.com/s/1qWLyCDu).
Simulation results using experimental data from the DIII-D tokamak demonstrate that the method is effective at driving the plasma profile to a predefined desired profile at the terminal time.
Nevertheless, there is still some room for improvement. For example, there are also some complicated state constraints (path constraints) in the realistic tokamak operation. We can also developed some effective computational methods embedded in our program to handle these conditions (see \cite{linsurvey2013,loxton2008optimal}). Some more accurate dynamic models for the prediction of the evolution of poloidal flux profile can also be further considered as well as steady-state operation for the flat-top phase in the plasma current evolution. Furthermore, the proposed framework of solving PDE-constrained optimization problem in this paper has also the prospective to solve the real general applications arising in the industrial fields.

\section*{References}
\bibliographystyle{plain}
\bibliography{RenXu1}

\end{CJK*}
\end{document}